\def\Journal#1#2#3#4{{#1} {\bf #2}, #3 (#4)}

\def\NPA{{ Nucl. Phys.}  A}

\def\PRL{ Phys. Rev. Lett.}
\def\PRC{{ Phys. Rev.} C}

\documentclass[
    ,final            
  ]
  {aipproc}

\layoutstyle{6x9}
\usepackage{epstopdf}

\begin{document}

\title{High Rapidity Physics with the BRAHMS Experiment}


\author{R. Debbe for the BRAHMS Collaboration}{
  address={Brookhaven National Laboratory}
}

\begin{abstract}
 We report the study of the nuclear modification factor $R_{AuAu}$ as function of $p_{T}$ and pseudo-rapidity in Au+Au collisions 
at top RHIC energy.
We find this quantity almost independent of pseudo-rapidity. We use the $\bar{p}/\pi^{-}$ ratio as a
 probe of the parton density and the degree of thermalization of the medium formed by the collision. The $\bar{p}/\pi^{-}$ ratio 
has a clear rapidity dependence. The combination of these two measurements suggests that the pseudo-rapidity dependence of the  
$R_{AuAu}$ results from the competing effects of energy loss in a dense and opaque 
medium and the modifications of the wave function of the high energy beams in the initial state.
\end{abstract}

\maketitle


\section{Introduction}

The BRAHMS Collaboration has completed a thorough program exploring the particle production from several nuclear 
systems and several energies at RHIC. The wide and almost continuous rapidity coverage of the BRAHMS spectrometers 
stands out in the RHIC program. This coverage gives us the ability to probe deeper into the beam wave functions, 
that is, to be sensitive  to the smallest values of the 
longitudinal momentum fraction carried by its partons. The particle spectra from A+A collisions have been 
compared to p+p yields measured at the same energy and with the same detectors, the yields from p+p collisions are 
scaled by the estimated number of binary collisions in the A+A systems $\langle N_{coll}\rangle$. The comparisons are done with the so
called Nuclear modification factor: $R_{AA} = Y^{AA}/\langle N_{coll}\rangle Y^{pp}$ where $Y^{AA}$ denotes the invariant yield 
extracted from A+A collisions, and $Y^{pp}$ the one from p+p collisions.  If the A+A system is an incoherent 
superposition of nucleon+nucleon collisions, the ratio should be constant and equal to one. The result of the 
measurements has been a dramatic suppression (by as much as a factor of 5 at the highest energy) that extends to 
high values of transverse momentum ($\sim 20 GeV/c$). All RHIC experiments have already reported these results at
mid-rapidity and they are now considered as a consequence of the formation of a dense and opaque medium dubbed sQGP 
\cite{WhitePapers}. Such a conclusion was reached after the nuclear modification factor extracted for d+Au collisions 
showed a Cronin type enhancement at mid-rapidity
implying that the suppression seen in Au+Au events is a final state effect. The study of d+Au collisions as function of rapidity
has generated renewed interest in the contribution to particle production in hadron-hadron interactions coming from partons with 
the smallest longitudinal momentum fraction $x$. 
The high rapidity suppression measured in d+Au colisions 
 \cite{RdA} has been described as the modification of an already saturated gluonic
system present at RHIC energies. Such modification  results from the interplay of additional gluon emission as well as gluon 
fusion in the initial state of the
interaction \cite{Wiedemann,KKT}. This modification of the nuclei wave function should also be present in A+A collisions
but may be masked by the fact that any measurement at high rapidity probes the projectile as well as the target 
fragmentation regions of both beams (here, for convenience, we borrow the naming scheme familiar in fixed target p+A physics).

\section{The $R_{AuAu}$ factor as a function of Rapidity}

\begin{figure}
  \includegraphics[height=.3\textheight]{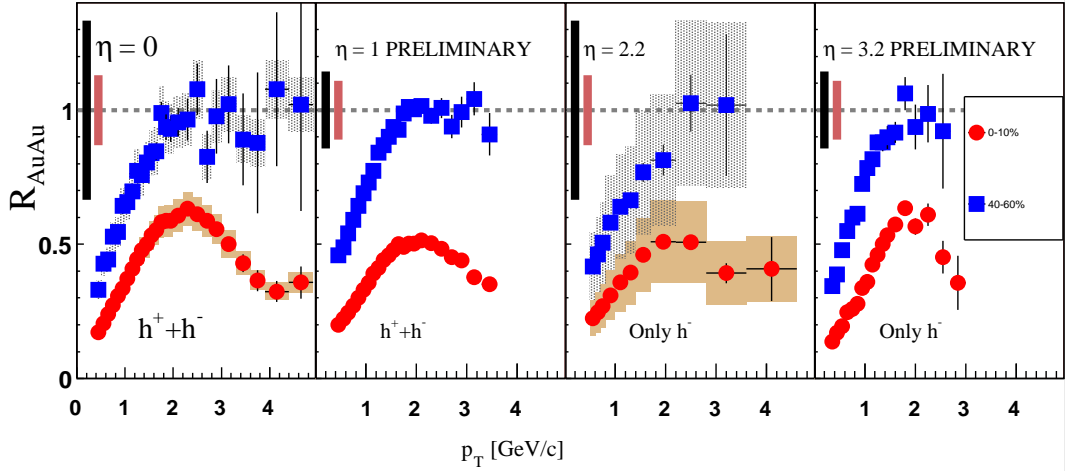}
  \caption{Nuclear modification factor $R_{AuAu}$ at $\sqrt{s_{NN}}=200$ GeV extracted from two centrality samples: central 0-10\% shown 
with filled circles (red online) and semi-central 40-60\% shown with filled squares (blue online), at four values
of pseudo-rapidity: $\eta=0$ and 1. for charged particles $h^{-}+h^{+}$, and $\eta=2.2$ and 3,2 for negative hadrons $h^{-}$.
 Statistical 
errors are shown with bars and the estimated systematic errors are shown with boxes, the normalization errors are
shown on the left of each panel and centered at the value of 1. }
  \label{fig:RaaRapidity}
\end{figure}

Figure  \ref{fig:RaaRapidity} shows the nuclear modification factor  extracted from Au+Au collisions at 
$\sqrt{s_{NN}}=200$ GeV/c versus $p_{T}$ and pseudo-rapidity $\eta$. Each panel shows the 
$R_{AuAu}$ factor for charged hadrons extracted from two centrality samples: The factors extracted from the most central 
events (0-10\%) are shown
with filled circles (red online), and the ones extracted from semi-central events (40-60\%) are shown with filled squares (blue online). 
At $\eta=2.2$ and $\eta=3.2$ the $R_{AuAu}$ factor was  calculated with negative hadrons.
The results at $\eta=0$ and 2.2 have been published and more details about the analysis can be found in \cite{PRL91}.

\begin{figure}
  \includegraphics[height=.21\textheight]{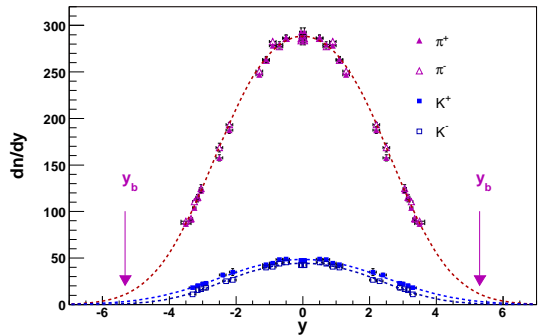}
  \caption{Rapidity density for charged pions and kaons produced in the 0-5\% central Au+Au collisions at $\sqrt{s_{NN}}=200$ GeV.}
  \label{fig:pionYield}
\end{figure}

Remarkably, Fig. \ref{fig:RaaRapidity} shows that the $R_{AuAu}$ reaches an almost constant  maximum at all rapidities 
ranging from 0 on the left-most panel to 3.2 on the right. Naively, if  these 
measurements were controlled by energy loss in the medium, 
one might expect less suppression at $\eta=3.2$ because the measured
pion yield shown in Fig. \ref{fig:pionYield} drops  by a factor of 2 between y=0 and y=3 (while $\langle p_{T} \rangle$ changes by
at most 10\%). If we make use of 
the parton-hadron duality hypothesis, this variation in the pion density will also be present in the gluon density
of the formed medium.

The absence of rapidity dependence in the $R_{AuAu}$ is a puzzle that may be difficult to solve because many effects 
compete in the transverse momentum range of the measurement. For a strong energy loss that results in the preferential detection
of particles that originate close to the surface of the medium, the softening of the spectra at high rapidity can produce a similar effect.
We propose in this work the use of the 
$\bar{p}/\pi^{-}$ ratio as a tool to
characterize the evolution of that system with rapidity; The comparison of the yields of anti-protons and  negative pions
shows a strong rapidity dependence that can be contrasted with the one found to be practically absent in the $R_{AuAu}$ ratio.

\section{The rapidity dependence of $\bar{p}/\pi^{-}$ ratio}

In $e^{+}e^{-}$ annihilations 
around the Z pole, the $\bar{p}/\pi^{-}$ ratio  
 is small and doesn't exceed 0.2 \cite{epluseminus}. This ratio is the result
of single parton fragmentation in the vacuum. Such a mechanism favors the production of many particles sharing the original
parton momentum. Because the parton is colored, an assumed string breaking mechanism will favor the production of mesons over that
of baryons \cite{recombination}.
At RHIC, the $\bar{p}/\pi^{-}$ 
and $p/\pi^{+}$ ratios have been
found to reach  values close to 1 and in some cases, even bigger than 1 at intermediate values of $p_{T}$ (2-3GeV/c). 
In a medium with high parton density, fragmentation and recombination compete in the formation of hadrons, and
recombination with its additive nature may be more efficient at forming particles at higher values of $p_{T}$. 
The same mechanism would also generate
more protons at intermediate $p_{T}$, since the momentum a protons would be the sum of momenta from three partons in 
contrast to the two
required to form a meson. These measured high values of the $\bar{p}/\pi^{-}$ and $p/\pi^{+}$ ratios  have been
explained as resulting from the  recombination of partons present in the thermalized medium 
 \cite{recombination}. 

\begin{figure}
  \includegraphics[height=.3\textheight]{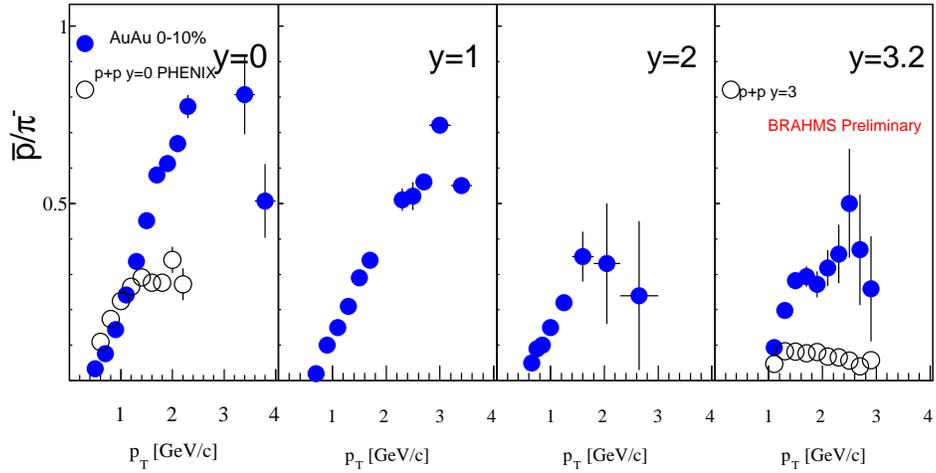}
  \caption{The $\bar{p}/\pi^{-}$ ratio extracted from Au+Au collisions at $\sqrt{s_{NN}}=200$ GeV as function of transverse momentum at
four values of rapidity. The ratios were extracted from central events 0-10\%. The same ratio extracted from 
p+p collisions is shown with open circles at y=0 and y=3.2. The errors are statistical.}
  \label{fig:rapidity}
\end{figure}

The $\bar{p}/\pi^{-}$ ratio extracted from Au+Au collisions at top RHIC energy is shown in Fig. \ref{fig:rapidity} with
filled circles. The left-most panel shows the ratio at mid-rapidity reaching values as high as 4 times the values
found for the so called fragmentation in the vacuum. As rapidity changes to y=1 the maximum of the ratio decreases
visibly and at the highest rapidity (y=3.2) on the right of the figure, the ratio reaches its smallest value. 
For comparison, the same ratio extracted from p+p collisions by the PHENIX collaboration at y=0 is shown in the 
left-most panels as well as the same ratio measured by 
BRAHMS  at y=3.2, also from p+p collisions, shown in the right-most panel.

\begin{figure}
  \includegraphics[height=.3\textheight]{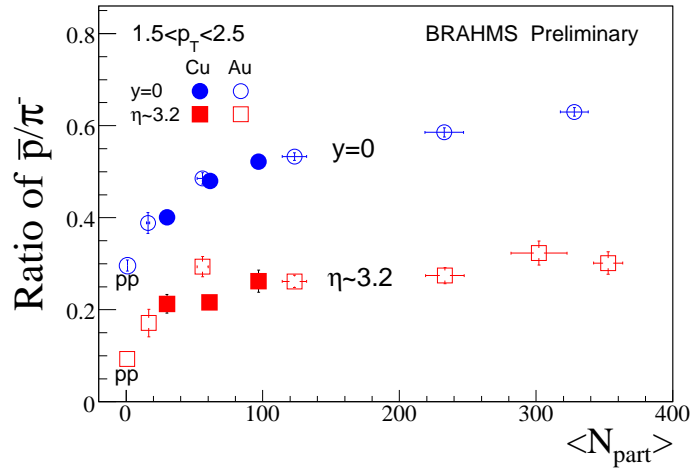}
  \caption{Compilation of the integral of the $\bar{p}/\pi^{-}$ over the $1< p_{T} < 2.5 $ GeV/c interval for various
colliding systems at $\sqrt{s_{NN}}=200$ GeV. The integral 
for Cu+Cu 
collisions at y=0 shown with filled circles (blue online), and y=3.2 with filled squares (red online). The results 
obtained from Au+Au collisions are shown with open circles (blue online) at y=0, and open squares (red online) at y=3.2.
The p+p results are shown with similar symbols and are located at $<N_{part}>=0$. }
   \label{fig:centrality}
\end{figure}

Figure \ref{fig:centrality} summarizes the variation of the $\bar{p}/\pi^{-}$ ratio with the calculated  mean number of 
participant nucleon $\langle N_{part}\rangle$ at a determined centrality of the 
collision. The integral over $p_{T}$ in an interval that includes the maxima seen in all panels of Fig. \ref{fig:rapidity},
should convey the evolution of this ratio in shape as well as magnitude as function of rapidity.  
For the purpose of this report, the main feature in Fig. \ref{fig:centrality} is the fact that the $\bar{p}/\pi^{-}$ ratio extracted from
 the most central Au+Au collisions 
at y=3 has the same value as the one measured at y=0 in to p+p collisions at the same energy.

 If recombination is the effect that drives the $\bar{p}/\pi^{-}$ ratio, its clear 
rapidity dependence seen in Fig. \ref{fig:rapidity} puts in doubt any explanation of the suppression measured with the
$R_{AuAu}$ factor at high rapidity as being solely produced by energy loss in a dense medium. The behavior of the 
$\bar{p}/\pi^{-}$ ratio
seems to indicate that the sQGP does not extend to high rapidity, or if it does, its effects on partons traversing it at high
rapidity cannot be strong.  The small values of the $R_{AuAu}$ factor at high rapidity may rather 
be due to other effects that stand out as  energy loss weakens as one approaches the beam fragmentation regions. 
One such effect at high rapidity may be the  modification of the beam wave functions  as they are
now probed into even smaller values of x.

\section{Summary}

We have shown the almost rapidity independent nuclear modification factor $R_{AuAu}$ extracted from Au+Au collisions at two 
centralities. The suppression of the $R_{AuAu}$ at mid-rapidity is produced by energy loss in the dense and opaque  sQGP   formed 
early in the Au+Au collisons. We postulated the use of the $\bar{p}/\pi^{-}$ to probe the density and degree
of thermalization of that medium at different rapidities. The strong rapidity dependence of the  $\bar{p}/\pi^{-}$ ratio led us to 
conclude that the effects of the dense and opaque sQGP do not extend to high rapidity.  The continued suppression seen in the
$R_{AuAu}$ factor may then be the result of a compromise between energy loss in the sQGP that dominates around mid-rapidity and
the modification of the beam wave functions that become apparent at high rapidity. 


The author is greatful to the Conference organizers for a truly exciting experience. 
This work was supported by 
the Office of Nuclear Physics of the Office of Science of the U.S. Department of Energy, 
the Danish Natural Science Research Council, 
the Research Council of Norway, 
the Polish State Committee for Scientific Research (KBN) 
and the Romanian Ministry of Research.

\end{document}